\documentstyle[aps,preprint]{revtex}
\input{epsf}
\begin{document}
\title{Arrested Cracks in Nonlinear Lattice Models of Brittle Fracture}
\author{David A. Kessler\cite{barilan}}
\address{Dept. of Mathematics, Lawrence Berkeley National Laboratory,
1 Cyclotron Road, Berkeley, CA 94720}
\author{Herbert Levine}
\address{Dept. of Physics,
University of California, San Diego La Jolla, CA  92093-0319}
\maketitle
\begin{abstract}
We generalize lattice models of brittle fracture to arbitrary nonlinear
force laws and study the existence of arrested semi-infinite cracks.
Unlike what is seen in the discontinuous case studied to date, the
range in driving displacement for which these arrested cracks exist is
either very small or precisely zero. Also, our results indicate that small
changes in the vicinity of the crack tip can have an extremely large effect
on arrested cracks. Finally, we briefly discuss the possible relevance of our 
findings to recent experiments.
\end{abstract}

Recent years have seen a rebirth of interest by the physics community
in the issue of dynamic fracture. This is due to a variety of new experimental
results which are not explainable within the confines of the traditional
engineering approach to fracture~\cite{freund}. These results include a dynamical
instability to micro-branching~\cite{texas,fineberg}, the formation of
non-smooth fracture surfaces~\cite{steinberg} and the rapid variation of the fracture
energy (including dissipative losses incurred during cleavage) with
crack velocity~\cite{energy}. These issues are reviewed in a recent paper by Fineberg
and Marder~\cite{frac-review}.

One approach for dealing with dynamic fracture involves restricting the
atomic interactions to those occurring between neighboring sites of an originally
unstrained lattice. These lattice models can never be as realistic as full
molecular dynamics simulations, but compensate for this shortcoming by
being much more amenable to analysis, both numerical and (via the Wiener-Hopf
technique) otherwise. This approach was pioneered by Slepyan and 
co-workers~\cite{slepyan} and further developed by Marder and Gross~\cite{marder}
and most recently by ourselves~\cite{kl-cracks}. Most of the results to
date have been obtained using a simplified force law which is linear
until some threshold displacement at which point it drops abruptly to zero.
Below, we will study a generalization for which the force is a smooth function
of the lattice strain. One of our goals is to learn which aspects of
fracture are sensitive to microscopic details and which are universal.

One interesting aspect of these lattice models concern the existence of a
range of driving displacements $\Delta$ for which non-moving semi-infinite
crack solutions can be found. For the aforementioned discontinuous force model,
there exists a wide range of these arrested cracks. For example, ref.~\cite{kl-cracks}
found that $\Delta$ could range from 40\% below to 40\% above the Griffith
displacement $\Delta _G$, the driving at which it first becomes energetically
favorable for the system to crack. This phenomena is 
connected to the existence of a velocity gap, i.e. a minimal velocity
for stable crack propagation. Experimentally, no such gap has been reported,
even for materials such as single-crystal silicon~\cite{texas-silicon} which 
should be at least approximately
describable by lattice models. It is therefore of some interest to study how
the arrested crack range depends on the microscopic details of the
assumed atomic force law. Here we present the results of such as a study,
including the finding that this range drops rapidly towards zero as the
force law is made smoother and hence more realistic.

As in ref.~\cite{kl-cracks}, we work with a square lattice and with 
scalar displacements (mode III). We focus on arrested cracks and
write the static equation as
\begin{equation}
0 \ = \   -f \left( u_{i+1,j}-u_{i,j} \right)
+ f \left( u_{i,j}-u_{i-1,j} \right) 
-f \left( u_{i,j+1}-u_{i,j} \right)
+ f \left( u_{i,j}-u_{i,j-1} \right)
\end{equation}
Here the indices $\{ i,j \} $ label the lattice site and $u$ is the displacement.
Sites on the last row of the the lattice, $j=N_y$, are coupled to a
row with fixed displacement $\Delta$. The first row, $j=1$, is coupled to
a $j=0$ displacement field $u_{i,0}$ which via symmetry equals $-u_{i,1}$.
Finally, $f$ is a nonlinear function of its argument, the lattice strain.
We investigate two forms~\cite{odd-note}:
\begin{equation}
f_e (u) \ = \ -u \frac{1+\tanh{(\alpha (1-u))}}{1+\tanh{\alpha }}
\end{equation}
\begin{equation}
f_p (u) \ = \ -\frac {u \alpha^{\alpha +1} }{(u+\alpha )^{\alpha +1} }
\end{equation}
For both of these forms, increasing $\alpha$ reduces the length scale over
which $f$ falls to zero once outside the Hooke's law regime ($u <1$). The
exponential force $f_e$ reduces to the familiar discontinuous force
(linear until complete failure) as $\alpha \rightarrow \infty$.

Our procedure for finding solutions is in principle straightforward. At
large positive $i$ in the uncracked material, we know that the system will
adopt a uniformly strained state. Conversely, at large negative $i$ the
cracked state will have a large displacement $u_{i,1}$ and (almost) zero
strains for $j>1$. Fixing the boundary condition $\Delta$ allows us to easily
find these asymptotic states. Once found, these solutions are used as
fixed displacements for the columns $i=N_x+1$ and $i=-N_x-1$ respectively.
The arrested crack then requires us to solve for $(2N_x+1)N_y$ variables.
We impose the equation at motion at all sites except for the crack ``tip",
($ i=0,j=1$) where instead we specify the displacement; this approach
perserves the banded structure of the system. Newton's algorithm
then allows us to converge to a solution. Afterwards, the residual equation
of motion becomes a solvability condition with which $\Delta$ can be
determined. The range of allowed values of $\Delta$ for arrested cracks is
found as one systematically sweeps through the value of the aforementioned
fixed displacement.

In Fig. 1 we present our results for the exponential model. For illustration,
we have chosen to show data for $N_y=10$  as a function of $\alpha$. For
large $\alpha$, the range of $\Delta$ is large and there is a marked
asymmetry between the rising segment of $\Delta$ versus imposed displacement
and the (much steeper) falling segment. As $\alpha \rightarrow \infty$,
the falling portion becomes vertical. These segments represent different
crack solutions at fixed $\Delta$; as $\Delta$ reaches the end of its allowed
range, these solution branches collide and disappear in a standard saddle-node
bifurcation point. To verify this, we have performed~\cite{stability} a linear stability calculation
of these solutions, assuming purely inertial dynamics (i.e. setting the left hand
side of Eq. 1 to $\ddot{u}_{i,j}$). As expected, there is a single mode 
of the spectrum for the growth rate $\omega$  for which $\omega ^2$
goes from negative to positive as we go up the rising segment, reach the
maximal driving, and then go back down.
 
\global\firstfigfalse
\begin{figure}
\centerline{\epsfxsize=4.25in \epsffile{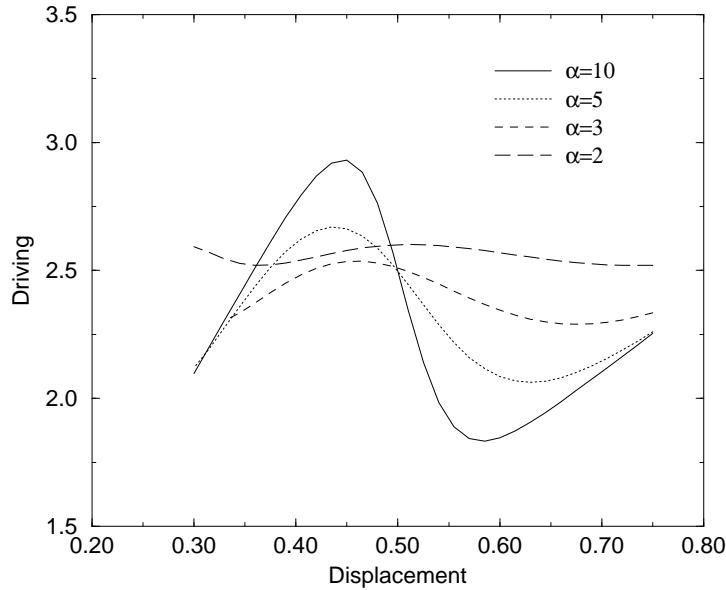}}
\caption{$\Delta$ versus imposed $u_{0,1}$ displacement for different
values of $\alpha$ in $f_e$. All data has $N_y=10$ and $N_x=100$.}
\label{fig1}
\end{figure}

Fig. 1 demonstrates that as the potential is made smoother, the range
of arrested cracks shrinks dramatically. In Fig. 2, we show this range as a
percentage of $\Delta _G$. The best fit to our data suggests that the
range vanishes as an essentially singular function of $\alpha$,
\begin{equation}
\frac{\Delta_{\mbox {max}} - \Delta_{\mbox {min}}}{\Delta _G} \ \sim \
A \exp {-\frac{\alpha _0}{\alpha }}
\end{equation}
where for $N_y =10$, $\alpha _0 \simeq 6.6$ and otherwise is a slowly
varying function of $N_y$ as long as the system is sufficiently large
compared to the potential fall-off. 

\begin{figure}
\centerline{\epsfxsize=4.25in \epsffile{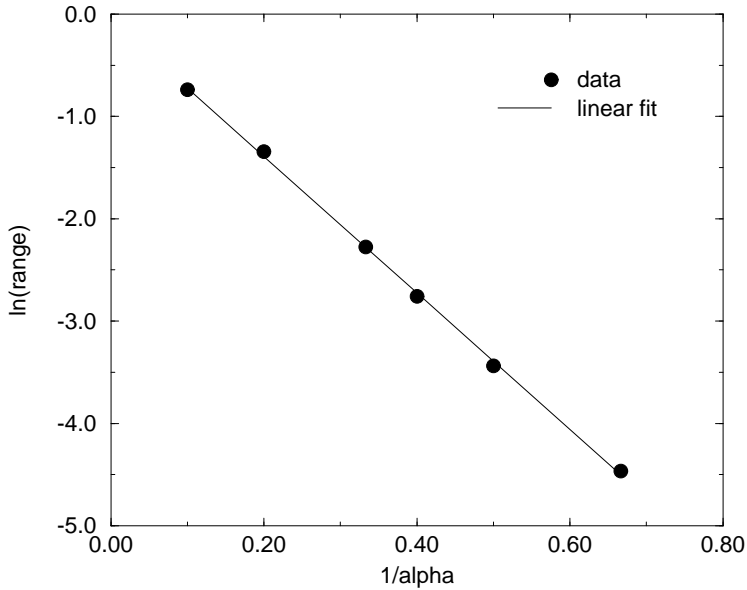}}
\caption{Arrested crack range normalized by the Griffith 
displacement $\Delta _G$ versus $\alpha$ in $f_e$ ;
again all data is for $N_y=10$, $N_x=100$. }
\label{fig2}
\end{figure}

Let us now turn to the power-law form. Based on our findings 
above, we would expect that this rather smooth force law would
give rise to a range which is practically zero.
We have verified this prediction in two ways. First, for the
case $\alpha =3$ we performed our usual scan
over imposed $u_{0,1}$ displacement and noted that the selected $\Delta$
varies by less than $10^{-6}$. Second, we computed the stability
spectrum and found a mode at $\omega ^2 < 10^{-6}$; this value is
indicative of how close we are at a randomly chosen displacement to the 
extremal value of $\Delta$ at the saddle-node bifurcation . These numbers
are consistent with our numerical accuracy and hence the true range is
probably even smaller. Needless to say, ranges of this size would be
unmeasurable. It is interesting to point out that the almost-zero mode
is nothing other than a spatial translation of the crack. That is,
translating the crack with respect to the underlying fixed lattice is
almost a symmetry of the solution. 

So, by making the potential smoother one tends to eliminate arrested crack
solutions. How does this change come about? To try to address this question,
we plot in Fig. 3 the lattice strain  field $u_{i+1,1}-u_{i,1}$ for $-N_x \leq i \geq N_x$
for the three potentials, exponential with $\alpha = 5$ or $2$ and power-law
with $\alpha =3$. For this comparison, we have found (stable) solutions with
$u_{0,1} =.75$ for all three potentials, and then normalized the
strains by dividing with the respective values of $\Delta$. First, we note
that beyond $x \simeq 5$, the different cases are virtually indistinguishable
and all lie on the expected $x^{-1/2}$ universal curve~\cite{freund,kl-cracks}.
The interior ``process-zone" region is affected by changing the potential,
but rather minimally. For example, the two exponential cases differ in
only one or two points, yet this is sufficient to shrink the arrested crack
range by almost an order of magnitude. The power-law choice has a process-zone
which is a bit wider and there is less maximal strain, but that is all.
We thus conclude that the existence and size of the arrested crack range
are extremely sensitive to microscopic details! We note in passing that the
process-zone for any specific potential quickly reaches an asymptotic size
once $N_y$ is sufficiently large and in particular does not increase
indefinitely in the macroscopic limit. 
Treatments~\cite{langer-recent,barenblatt} which include a
mesoscopic size ``cohesive-zone" are therefore not accurate representations
of this class of lattice models.

In a recent experiment~\cite{texas-silicon} on fracture in silicon, no arrested cracks
were observed. A molecular dynamics simulation using a modified 
Stillinger-Weber potential also exhibited no arrested cracks 
when studied at high enough temperature. However, the potentials used
here were rather short-ranged, as compared with some estimates that
arise from density-functional theory~\cite{rucker}. Our results indicate
that increasing the range and thereby using smoother potentials will
eliminate (at least as far as experimentally attainable
precision is occurred) arrested cracks and may offer a simpler explanation
of the experimental finding than one which requires thermal creep. This
could of course be tested in principle by re-doing the experiments at a reduced
temperature.

\begin{figure}
\centerline{\epsfxsize=4.25in \epsffile{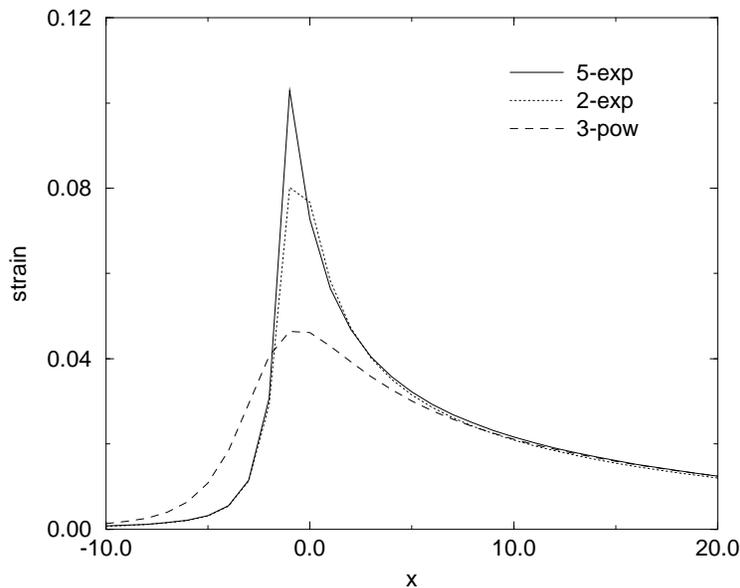}}
\caption{Strain $u_{i+1,1}-u_{i,1}$ for three different potentials. Data
is for $N_y=40$, $N_x =200$ and is normalized to the respective $\Delta$ 
values. The power-law curve has been shifted two sites to the left
so as to better match the field at large positive $i$.}
\label{fig3}
\end{figure}

\acknowledgments
HL acknowledges the support of the US NSF under grant DMR98-5735; DAK 
acknowledges the support of the Israel Science Foundation and the
hospitality of the Lawrence Berkeley National Laboratory. The work
of DAK was also supported in part by the Office of Energy Research,
Office of Computational  and Technology Research, Mathematical, Information 
and Computational Sciences Division, Applied Mathematical Sciences Subprogram,
of the U.S. Department of Energy, under Contract No. DE-AC03-76SF00098.
Also, DAK acknowledges useful conversation with  M. Marder and G. Barenblatt.

\end{document}